\documentstyle[twocolumn,prb,aps]{revtex}

\newcommand{\be}{\begin{equation}}
\newcommand{\ee}{\end{equation}}
\newcommand{\ba}{\begin{eqnarray}}
\newcommand{\ea}{\end{eqnarray}}

\begin{document}

\draft
\title{Anomalous dimensions of gauge-invariant amplitudes in massless
effective gauge theories of strongly correlated systems}

\author{V.~P. Gusynin}

\address{Institute of Theoretical Physics, Sidlerstrasse 5, CH-3012,
Bern, Switzerland\\
Bogolyubov Institute for Theoretical Physics,  Kiev-143, 03143,
Ukraine}

\author{D.~V. Khveshchenko}
\address{Department of Physics and Astronomy, University of North
Carolina, Chapel Hill, NC 27599, USA}

\author{M. Reenders}
\address{Department of Polymer Chemistry and Materials Science Center,\\
University of Groningen, Nijenborgh 4, 9747 AG Groningen, The
Netherlands}

\date{\today}

\maketitle

\begin{abstract}
We use the radial gauge to calculate the recently proposed ansatz
for the physical electron propagator in such effective models of
strongly correlated electron systems as the $QED_3$ theory of the
pseudogap phase of the cuprates. The results of our analysis help
to settle the recent dispute about the sign and the magnitude of
the anomalous dimension which characterizes the gauge invariant
amplitude in question and set the stage for computing other, more
physically relevant, ones.
\end{abstract}

\section{Introduction}
As a generic property, the one-dimensional Fermi systems with
short-range (screened) repulsive interactions routinely
demonstrate algebraic decay of all the correlation functions
governed by non-universal (coupling-dependent) anomalous
exponents.

A possibility of the emergence of a similar behavior, commonly
referred to as the "Luttinger liquid", in higher dimensional
strongly correlated electron systems has been extensively
discussed in the recent literature.

Thus far, however, no solid consensus has been reached even on the
necessary criteria that have to be fulfilled for the Luttinger
behavior to set in, much less on whether or not it occurs in any
specific example of a strongly correlated electron system. It was
largely for this reason that the attention has recently been drawn
to the class of effective models described by (possibly, spatially
anisotropic and/or Lorentz non-invariant) deformations of the
standard action of Quantum Electrodynamics.

Motivated by the puzzling properties of the quasi-two-dimensional
high-temperature copper-oxide superconductors, most of the
interest has been focused on the three-dimensional case described
by either the ordinary (parity-even) $QED_3$ or the abelian 3D
Chern-Simons theory \cite{Affleck-Marston,Mavromatos}, where the
finite density problem of non-relativistic (massive) fermions has
become the main subject of the scrutiny. However, the latter was
found to fall into a rather different class of non-Fermi liquids
which bear little resemblance to the 1D Luttinger liquid
\cite{Reizer}.

Recently, the idea of the conjectured Luttinger-like behavior has
been rekindled in the recent theories of the pseudogap phase of
the underdoped superconducting cuprates
\cite{Lee,Wen1,Franz1,Ye1,Herbut1}. Albeit describing rather
different physics, all these approaches resort to the same
effective description in terms of the pseudo-relativistic $QED_3$
theory of the gapless nodal fermion excitations which retain their
$d$-wave symmetrical gap well above the critical temperature $T_c$
regarded merely as the onset of global phase coherence.

Above $T_c$ the fermionic excitations experience strong scattering
by both thermal and quantum fluctuations of an incipient ordering,
such as a flux of the gauge field measuring a local spin chirality
\cite{Lee,Wen1} or vortex-antivortex pairs of the pairing order
parameter \cite{Franz1,Ye1,Herbut1}. The latter scenario has
recently received a new experimental support from the observation
that the vortex matter is present at temperatures well in excess
of $T_c$, as revealed by the measurements of the Nernst effect
\cite{Ong}.

As another important ingredient, the $QED_3$ theory of the
pseudogap phase was aimed at explaining the ubiquitous destruction
of the coherent quasiparticles above $T_c$ which was observed in
angular-resolved photoemission and tunneling experiments.

To this end, the authors of Ref.~\cite{Wen1} conjectured that the
electron propagator in question may, in fact, exhibit the
Luttinger behavior

\begin{equation}
G(x)\propto {\hat x}/|x|^{3+\eta}, \quad {\hat x}=\gamma_\mu
x_\mu, \label{propagator}
\end{equation}
characterized by a positive anomalous exponent $\eta>0$, and they
also attempted to fit the ARPES data, while claiming a good
agreement with experiment (unless explicitly stated otherwise,
throughout this paper we use the notation $\hat{n}\equiv
\gamma_\mu n_\mu$ for vectors $n_\mu$ contracted with the Dirac
matrices $\gamma_\mu$).

The conclusions drawn in Ref.~\cite{Wen1} were based on the use of
the following heuristic form of the gauge-invariant electron
propagator
\begin{equation}
G(x-y)=\langle 0|\psi(x) \exp{\left[-i\int \limits_\Gamma dz^\mu\,
A_\mu(z)\right]} \bar\psi(y)|0 \rangle, \label{Ginvdef}
\end{equation}
where the line integral was taken along the contour $\Gamma$
chosen as the straight-line segment connecting the end points.

Later on, the calculations of Ref.~\cite{Wen1} were carried out by
a number of other authors, and the results for the anomalous
exponent appeared to vary not only between the different authors
($\eta=32/3\pi^2N$\, \cite{Wen1} vs $-32/3\pi^2N$\, \cite{DVK1})
but also from one to another work of the same authors
($\eta=-16/3\pi^2N$\, \cite{Franz1} vs $16/3\pi^2N$\,
\cite{Franz2} and $\eta=32/3\pi^2N$\, \cite{Ye1} vs
$-64/3\pi^2N$\, \cite{Ye2}).

While some of the calculations were performed in the conventional
covariant gauges \cite{DVK1,Ye2}, other authors made use of the
potentially problematic axial gauge ($(x-y)_\mu A_\mu(z)=0$ where
$x,y$ are arbitrarily chosen points which are taken to coincide
with the end points of the contour $\Gamma$)
\cite{Wen1,Franz1,DVK1} which spurred a debate over the issue of a
true (vs limited, see \cite{DVK2}) gauge invariance of
Eq.~(\ref{Ginvdef}), as opposed to its surrogate functions
proposed in \cite{Franz2} (see Summary for a more extended
discussion).

While seemingly being an issue of secondary importance, a proper
construction of the physical electron propagator is, in fact,
imperative, as far as ascertaining the status of the conjectured
Luttinger-like behavior in the $QED$-like theories is concerned.

In light of the present controversy, in this paper we undertake
yet another attempt to settle the dispute about the physically
motivated form of the electron propagator and the actual value of
$\eta$ (if any) by resorting to the so-called radial
(Fock-Schwinger) gauge ($(z-x)_\mu A_\mu(z)=0$ and $x$ is an
arbitrary fixed point). The radial gauge is known to be free of
the potential problems that might exist in the axial gauge, which,
according to some authors, may even require one to introduce ghost
fields \cite{ghost-in-axial-gauge}. In addition, we also set out
to explore the dependence of the previously conjectured form of
the electron propagator (2) on the choice of the contour $\Gamma$.

\section{Gauge invariant fermion propagator in
the Fock-Schwinger gauge}
We start with the $3D$ relativistic
theory of massless Dirac spinors coupled to a massless $U(1)$
gauge field, whose Euclidean action is
\begin{eqnarray}
S\left[\psi,\bar\psi,A\right]=\int
d^3x\sum_{i=1}^N\bar\psi_i\gamma_\mu
\left(\partial_\mu-iA_\mu\right)\psi_i,
\end{eqnarray}
where $\bar\psi\equiv\psi^\dagger\gamma_0$ and the $N$-flavored
Dirac fermions are described by four-component bi-spinors which
belong to a reducible representation of the $\gamma$-matrices
satisfying the Clifford algebra $\{\gamma_\mu,
\gamma_\nu\}=2\delta_{\mu\nu}$ ($\mu,\nu=0,1,2$). The latter  can
be chosen in the form of the direct product $\gamma_\mu
=\sigma_3\otimes (\sigma_3,\sigma_2,\sigma_1)$ of the standard
triplet of the Pauli matrices $\sigma_\mu$.

In all of the above mentioned condensed matter-inspired QED-like
models the number of fermion flavors $N=2$. Nevertheless, in what
follows we choose to treat $N$ as a parameter that can assume
arbitrary values, depending on the problem in question.

The dynamics of the $U(1)$ gauge field is generated by the effective
action obtained after tracing out the
fermionic degrees of freedom
\begin{equation}
S_{\rm eff}[A]=\frac{1}{2}\int d^nx\int d^ny A_\mu(x)D^{-1}_{\mu\nu}(x-y)
A_\nu(y)+\dots,
\label{eff-gauge-action}
\end{equation}
where the dots stand for the higher order terms (non-Gaussian)
which we hereafter neglectm as is done in all of the previous
works on the subject. Intending to subsequently use the method of
dimensional regularization for evaluating Feynman diagrams we
formulated Eq.\ref{eff-gauge-action} in $n=3-\epsilon$ dimensions.
Also, in Eq.\ref{eff-gauge-action} we neglected the bare Maxwell
term $\sim (\partial_\mu A_\nu -\partial_\nu A_\mu)^2$ which turns
out to be irrelevant in the low-energy, long-distance limit.

The previously proposed candidate for the physical electron
propagator (2) studied in \cite{Wen1,Franz1,Ye1,DVK1,Franz2,Ye2}
can be cast in the following form
\begin{eqnarray}
G_{inv}(x-y) =\langle G[x,y;A]\exp{\left[-i\int
\limits_y^xdz^\mu\, A_\mu(z)\right]}\rangle, \label{line-integral}
\end{eqnarray}
where $G[x,y;A]$ is a fermion propagator for a given fixed
configuration of the gauge field $A_\mu$, and the brackets stand
for the (normalized) functional average over the gauge field which
is described by the action Eq.(\ref{eff-gauge-action}).

In the Euclidean momentum space, the kernel $D^{-1}_{\mu\nu}$ of
the quadratic operator has the form
\begin{equation}
D^{-1}_{\mu\nu}(q)=\frac{N\sqrt{q^2}}{8}\left(\delta_{\mu\nu}-\frac{q_\mu
q_\nu}{q^2}\right). \label{gauge-prop:q-space}
\end{equation}
Introducing a source field
\begin{equation}
J^\mu(z)=(x-y)^\mu\int\limits_0^1d\alpha\,\delta^n(z-y-(x-y)\alpha),
\label{Jdef}
\end{equation}
we can write the straight-line integral which appears in
Eq.~(\ref{Ginvdef}) as
\begin{equation}
\int \limits_y^x dz^\mu\, A_\mu(z)= \int d^nz\, J^\mu(z) A^\mu(z).
\label{JA}
\end{equation}
In the Fock-Schwinger gauge
\begin{equation}
(x-x_0)_\mu A_\mu(x)=0 \label{FS-gauge}
\end{equation}
the line integral in Eq.~(\ref{line-integral}) vanishes if one
chooses the reference point $x_0$ at the "center of mass"
$x_0=X=(x+y)/2$ (for the proof, see, Appendix \ref{FS-propagator}
which also contains a derivation of the photon propagator in the
FS gauge $D^{FS}_{\mu\nu}(z_1+X,z_2+X)$).

We compute the first order $1/N$ correction to the fermion
propagator (3) by expanding the inverse Dirac operator
$1/(\hat{\partial}-i\hat{A})=
1/\hat{\partial}+(1/\hat{\partial})(i\hat{A})(1/\hat{\partial})+\dots$
in powers of $A_\mu(z)$, thus obtaining
\begin{equation}
G_{inv}(x,y)=S(x-y)+\delta G(x,y),
\end{equation}
where $S(x-y)$ and $\delta G(x,y)$ are the bare fermion propagator
and the correction to it, respectively. The latter is given by the
expression
\begin{eqnarray}
&&\delta G(x,y)=
-\int d^n z_1 \int d^n z_2\, S(x-z_1-X)\gamma^\mu\nonumber\\
&&\times S(z_1-z_2)\gamma^\nu S(z_2+X-y)
D^{FS}_{\mu\nu}(z_1+X,z_2+X),
\end{eqnarray}
where it is natural to decompose the propagator $D^{FS}_{\mu\nu}$
for the gauge fields in the Fock-Schwinger gauge into four
contributions,
\begin{eqnarray}
D^{FS}_{\mu\nu}(x,y)=\sum_{i=1}^4D^{(i)}_{\mu\nu}(x,y),
\end{eqnarray}
as is done explicitly in Eqs.~(\ref{D1def}) to (\ref{D4def}). By
choosing the reference point in the Fock-Schwinger gauge to be the
center of mass $X$, one verifies by inspection of
Eqs.~(\ref{D1def}) to (\ref{D4def}) that the contribution $\delta
G$ can only depend on the relative coordinate $\bar x=x-y$:
\begin{eqnarray}
&&\delta G(\bar x)=-\int d^n z_1 \int d^n z_2\, S(\bar
x/2-z_1)\gamma^\mu
\nonumber\\
&&\times S(z_1-z_2)\gamma^\nu S(z_2+\bar
x/2)D^{FS}_{\mu\nu}(z_1+X,z_2+X). \label{gbar}
\end{eqnarray}
Taking the Fourier transform of Eq.~(\ref{gbar}) $FT[\delta G(\bar
x)]=i\delta G(p)$, $FT[S(\bar x)]=iS(p)$, $S(p)=1/\hat{p}$, we
obtain
\begin{eqnarray}
&&\delta G^{(i)}(p)=\int\frac{d^n q_1}{(2\pi)^n}\frac{d^n
q_2}{(2\pi)^n}\,
S(p+q_1/2+q_2/2)\gamma^\mu\nonumber\\
&&\times S(p-q_1/2+q_2/2)\gamma^\nu S(p-q_1/2-q_2/2)
D^{(i)}_{\mu\nu}(q_1,q_2).\nonumber\\
\label{Gidef}
\end{eqnarray}
In the above equation, the index $i=1,\dots,4$ labels the Fourier
transforms $D^{(i)}_{\mu\nu}(q_1,q_2)$ of the components
$D^{(i)}_{\mu\nu}(z_1+X,z_2+X)$ given by
Eqs.(\ref{D1def})-(\ref{D4def})
\begin{equation}
D^{(1)}_{\mu\nu}(q_1,q_2)=(2\pi)^n
\delta^n(q_1+q_2)\delta_{\mu\nu} D(q_1), \label{D1qdef}
\end{equation}
\begin{equation}
D^{(2)}_{\mu\nu}(q_1,q_2)=\int\limits_0^1d\alpha\,
q_{1\mu}\frac{\partial}{\partial q_{1\nu}} D^{(\alpha)}(q_1,q_2),
\label{D2qdef}
\end{equation}
\begin{equation}
D^{(3)}_{\mu\nu}(q_1,q_2)=\int\limits_0^1d\beta\,
q_{2\nu}\frac{\partial}{\partial q_{2\mu}} D^{(\beta)}(q_2,q_1),
\label{D3qdef}
\end{equation}
\begin{equation}
D^{(4)}_{\mu\nu}(q_1,q_2)=\int\limits_0^1\hspace{-0.8mm}d\alpha\int\limits_0^1\hspace{-0.8mm}
d\beta q_{1\mu}q_{2\nu} \frac{\partial}{\partial
q_{1\lambda}}\frac{\partial}{\partial q_{2\lambda}}
D^{(\alpha,\beta)}(q_1,q_2),\label{D4qdef}
\end{equation}
where
\begin{eqnarray}
D(q)&=&\frac{8}{N}\frac{1}{\sqrt{q^2}},\label{Ddef}\\
D^{(\alpha)}(q_1,q_2)&=&(2\pi)^n \delta^n(q_1+\alpha q_2) D(q_1/\alpha),\\
D^{(\alpha,\beta)}(q_1,q_2)&=&(2\pi)^n \delta^n(\beta q_1+ \alpha
q_2) D(q_1/\alpha).
\end{eqnarray}
The first contribution to the anomalous dimension stems from
$\delta G^{(1)}(p)$ given by Eqs.~(\ref{Gidef}) and
(\ref{D1qdef}). Since the Fourier transform of the propagator
Eq.(\ref{propagator}) is proportional to $\hat p/p^{2-\eta}$ it
proves convenient to multiply $\delta G^{(1)}(p)$ by $\hat p$ in
order to deduce the exponent $\eta$ which we are after. Now taking
the trace we arrive at the following result
\begin{eqnarray}
&&\frac{1}{4}{\rm Tr}\left[\hat p \delta
G^{(1)}(p)\right]=\frac{1}{4}
\int\frac{d^nq}{(2\pi)^n}\, {\rm Tr}[\hat p S(p)\gamma^\mu S(p-q)\gamma^\mu\nonumber\\
&&\times S(p)]D(q) =-\frac{8(n-2)}{N}\int \frac{d^n q}{(2\pi)^n}\,
\frac{p^2-p\cdot q}{\sqrt{q^2}p^2 (p-q)^2}\nonumber\\
&&=-\frac{4}{3\pi^2N}
\frac{|p|^{-\epsilon}}{\epsilon},\label{G1finres}
\end{eqnarray}
where we used dimensional regularization near $n=3-\epsilon$ in
the divergent integral
\begin{equation}
\int \frac{d^n q}{(2\pi)^n}\, \frac{p^2-p\cdot q}{\sqrt{q^2}p^2
(p-q)^2} =\frac{1}{6\pi^2} \frac{|p|^{-\epsilon}}{\epsilon}.
\label{feynmc1}
\end{equation}
Notably, Eq.~(\ref{G1finres}) coincides with the result for the
anomalous dimension of the ordinary (gauge-variant) fermion
propagator performed in the covariant Feynman gauge.

Next, we compute the term $\delta G^{(2)}(p)$ given by
Eqs.~(\ref{Gidef}) and (\ref{D2qdef}))
\begin{eqnarray}
&&\delta G^{(2)}(p)=\int\limits_0^1d\alpha \int\frac{d^n
q_1}{(2\pi)^n}\frac{d^n q_2}{(2\pi)^n}S(p+q_1/2+q_2/2)
\nonumber\\
&&\times\gamma^\mu S(p-q_1/2+q_2/2)
\gamma^\nu S(p-q_1/2-q_2/2) \nonumber\\
&&\times q_{1\mu}\frac{\partial}{\partial q_{1\nu}} (2\pi)^n
\delta^n(q_1+\alpha q_2) D(q_1/\alpha).
\end{eqnarray}
Integrating by parts we cast $\delta G^{(2)}(p)$ in the form
\begin{eqnarray}
&&\delta G^{(2)}(p)=-\int\limits_0^1d\alpha\int\frac{d^n
q_1}{(2\pi)^n}
\frac{d^n q_2}{(2\pi)^n}(2\pi)^n \delta^n(q_1+q_2\alpha)\nonumber\\
&&\times D(\frac{q_1}{\alpha})\frac{\partial}{\partial
q_{1\nu}}\left[
\left[S(p-\frac{q_1}{2}+\frac{q_2}{2})-S(p+\frac{q_1}{2}+\frac{q_2}{2})\right]\right.\nonumber\\
&&\times\left.\gamma^\nu S(p-\frac{q_1}{2}-\frac{q_2}{2})\right],
\label{g2mom1}
\end{eqnarray}
where we made use of the Ward-Takahashi identity (WTI) for the
bare propagators
\begin{equation}
S(k+q)\hat q S(k)=S(k)-S(k+q). \label{wti1}
\end{equation}
Multiplying Eq.~(\ref{g2mom1}) by $\hat p$ and taking the trace we
obtain
\begin{eqnarray}
&&\frac{1}{4}{\rm Tr}\left[ \hat p \delta G^{(2)}(p)\right]
=-\int\limits_0^1d\alpha\int\frac{d^n q_1}{(2\pi)^n}\frac{d^n q_2}{(2\pi)^n}\nonumber\\
&&\times(2\pi)^n \delta^n(q_1+q_2\alpha) D(q_1/\alpha)
\nonumber\\
&&\times \Biggr\{-\frac{(n-2)}{2}\frac{(2p^2-p\cdot q_1)}{
(p-q_1/2+q_2/2)^2 (p-q_1/2-q_2/2)^2}\nonumber\\
&&+ \frac{(n-2)}{2}\frac{(p\cdot q_1+p\cdot
q_2)}{(p-q_1/2-q_2/2)^2 (p+q_1/2+q_2/2)^2 } \Biggr\}.
\label{trhatpg2}
\end{eqnarray}
Being an odd function of the momenta $q_{1,2}$ the last term in
Eq.~(\ref{trhatpg2}) vanishes. Thus, we obtain
\begin{eqnarray}
&&\frac{1}{4}{\rm Tr}\left[ \hat p \delta G^{(2)}(p)\right]
=\frac{4(n-2) 2^{n}}{N}\int\limits_0^1d\alpha\int\frac{d^nq}{(2\pi)^n}\nonumber\\
&&\times\frac{(p^2+p\cdot q\alpha)}{\sqrt{q^2}(p+q(1+\alpha))^2
(p-q(1-\alpha))^2}.
\end{eqnarray}
The momentum integral can be evaluated by virtue of the Feynman
parameterization
\begin{eqnarray}
&&I_1=\int\frac{d^nq}{(2\pi)^n}\, \left[\frac{(p^2+p\cdot
q\alpha)}{\sqrt{q^2}(p+q(1+\alpha))^2 (p-q(1-\alpha))^2}\right]
\nonumber\\
&&=\frac{p^2}{(1-\alpha^2)^3}\frac{3}{4}\int\limits_0^1
dx\int\limits_0^1 dy\,
\frac{x}{\sqrt{1-x}}(1-(2 xy-x) \alpha\nonumber\\
&&-(1-x)\alpha^2)\int\limits_0^\infty \frac{dq}{2\pi^2}\frac{q^{n-1}}{[q^2+c]^{5/2}}\nonumber\\
&&=\frac{p^2}{(1-\alpha^2)^3}\frac{3}{4}\int\limits_0^1
dx\int\limits_0^1 dy\,\frac{x}{\sqrt{1-x}}(1-(2 xy-x) \alpha\nonumber\\
&&-(1-x)\alpha^2)\frac{c^{(n-5)/2}}{6\pi^2},
\end{eqnarray}
where the argument $c$ takes the form
\begin{eqnarray}
c&=&\frac{p^2 x}{(1-\alpha^2)^2}\left[1-x(1-2y)^2+2(1-x)(1-2y)\alpha\right.\nonumber\\
&+&\left.(1-x)\alpha^2\right].
\end{eqnarray}
The leading divergence of the integral over $\alpha$ can be
extracted by making an approximation similar to that of
Ref.~\cite{st84}
\begin{eqnarray}
&&c^{(n-5)/2}\approx \left[\frac{p^2}{(1-\alpha^2)^2}\right]^{(n-5)/2}\nonumber\\
&&\times\frac{1}{x\left[1-x(1-2y)^2+2(1-x)(1-2y)\alpha+(1-x)\alpha^2\right]}.\nonumber\\
\end{eqnarray}
In this way, we obtain
\begin{eqnarray}
I_1\approx \frac{1}{8\pi^2(1-\alpha^2)}
\left[\frac{|p|}{(1-\alpha^2)}\right]^{-\epsilon} I(\alpha),
\end{eqnarray}
where
\begin{eqnarray}
&&I(\alpha)=\int\limits_0^1 dx\int\limits_0^1 dy\nonumber\\
&&\times\frac{(1-x)^{-1/2}(1-(2 xy-x) \alpha-(1-x)\alpha^2)}{
\left[1-x(1-2y)^2+2(1-x)(1-2y)\alpha+(1-x)\alpha^2\right]}.\nonumber\\
\end{eqnarray}
The expression for $\delta G^{(2)}$ now reads
\begin{eqnarray}
\frac{1}{4}{\rm Tr}\left[ \hat p \delta G^{(2)}(p)\right]= \frac{4
|p|^{-\epsilon}}{\pi^2 N}\int\limits_0^1
d\alpha\,\frac{I(\alpha)}{(1-\alpha^2)^{1-\epsilon}}\nonumber\\
\approx
I(1)\int\limits_0^1\hspace{-0.8mm}d\alpha\frac{1}{2(1-\alpha)^{1-\epsilon}}
=\frac{4}{N\pi^2} \frac{|p|^{-\epsilon}}{\epsilon},
\label{G2finres}
\end{eqnarray}
where we used the integral
\begin{eqnarray}
I(1)&=&\int\limits_0^1dx\int\limits_0^1dy\,
\frac{x(1-x)^{-1/2}}{2[1+x(y-1)]}=2.
\end{eqnarray}
It can be readily shown that Eq.~(\ref{G2finres}) is also
identical to the result for ${\rm Tr}[\hat p \delta G^{(3)}(p)]$,
given by Eqs.~(\ref{Gidef}) and (\ref{D3qdef}).

Lastly, the expression for $\delta G^{(4)}(p)$ given by
Eqs.~(\ref{Gidef}) and (\ref{D4qdef}) reads as
\begin{eqnarray}
&&\delta G^{(4)}(p)=\int\limits_0^1 \frac{d\alpha}{\alpha^{n-2}}
\int\limits_0^{1/\alpha} d\tau
\int\frac{d^n q_1}{(2\pi)^n}\frac{d^n q_2}{(2\pi)^n}\nonumber\\
&&\times(2\pi)^n \delta^n(q_1 \tau +q_2)
D(q_1)\frac{\partial}{\partial q_1^\lambda}
\frac{\partial}{\partial q_2^\lambda}
\left[S(p+\frac{q_1}{2}+\frac{q_2}{2})\right.
\nonumber\\
&&\times\left.\hat q_1 S(p-\frac{q_1}{2}+\frac{q_2}{2})\hat q_2
S(p-\frac{q_1}{2}-\frac{q_2}{2})\right],\nonumber\\
\label{G4}
\end{eqnarray}
where $\tau=\beta/\alpha$. First, we make use of the WTI given by
Eq.~(\ref{wti1}) to simplify the product of the free fermion
propagators
\begin{eqnarray}
&&S(p+\frac{q_1}{2}+\frac{q_2}{2})\hat q_1 S(p-\frac{q_1}{2}+
\frac{q_2}{2})\hat q_2 S(p-\frac{q_1}{2}
-\frac{q_2}{2})\nonumber\\
&&=S(p+\frac{q_1}{2}+\frac{q_2}{2})-S(p-\frac{q_1}{2}+\frac{q_2}{2})\nonumber\\
&&+S(p+\frac{q_1}{2}+\frac{q_2}{2})\hat q_1
S(p-\frac{q_1}{2}-\frac{q_2}{2}).\label{doublewti}
\end{eqnarray}
As a result, Eq.~(\ref{G4}) assumes the following form
\begin{eqnarray}
&&\delta G^{(4)}(p)=\int\limits_0^1
\frac{d\alpha}{\alpha^{n-2}}\int\limits_0^{1/\alpha} d\tau
\int\frac{d^n q_1}{(2\pi)^n}\frac{d^n q_2}{(2\pi)^n}\nonumber\\
&&\times(2\pi)^n \delta^n(q_1 \tau +q_2)D(q_1)\nonumber\\
&&\times\frac{\partial}{\partial q_1^\lambda}
\frac{\partial}{\partial q_2^\lambda}
\left[S(p+\frac{q_1}{2}+\frac{q_2}{2})-S(p-\frac{q_1}{2}+\frac{q_2}{2})\right],\nonumber\\
\label{eqG4wqdef}
\end{eqnarray}
where we have dropped the last term in Eq.~(\ref{doublewti}) which
vanishes upon angular integration. Next, we pull the derivatives
to the front of the integral
\begin{eqnarray}
&&\delta G^{(4)}(p)=\frac{1}{4}\frac{\partial }{\partial
p_\lambda} \frac{\partial }{\partial p_\lambda} \int\limits_0^1
\frac{d\alpha}{\alpha^{n-2}} \int\limits_0^{1/\alpha}
d\tau\int\frac{d^n q_1}{(2\pi)^n}\frac{d^n
q_2}{(2\pi)^n}\nonumber\\
&&\times(2\pi)^n \delta^n(q_1 \tau +q_2)D(q_1)
\left[S(p+q_1/2+q_2/2)\right.\nonumber\\
&&+\left.S(p-q_1/2+q_2/2)\right] \label{eqG4wpdef}
\end{eqnarray}
and carry out the momentum integration over $q_2$ followed by
rescaling of the remaining momentum variable $q_1$ which yields
\begin{eqnarray}
\delta G^{(4)}(p)&=&\frac{1}{4} \frac{\partial }{\partial
p_\lambda} \frac{\partial }{\partial p_\lambda} \int\limits_0^1
\frac{d\alpha}{\alpha^{n-2}} \int\limits_0^{1/\alpha} d\tau\,
\int\frac{d^n q}{(2\pi)^n}\,D(q)\nonumber\\
&\times& \left[S(p+q(1-\tau)/2)+S(p-q(1+\tau)/2)
\right]\nonumber\\
&=& \frac{2^{n}}{N} \frac{\partial }{\partial p_\lambda}
\frac{\partial }{\partial p_\lambda} \int\limits_0^1
\frac{d\alpha}{\alpha^{n-2}} \int\limits_0^{1/\alpha} d\tau\,
\left[(1-\tau)^{1-n}\right.\nonumber\\
&+&\left.(1+\tau)^{1-n}\right]
\int\frac{d^n q}{(2\pi)^n}\,\frac{(\hat p+\hat q)}{\sqrt{q^2}(p+q)^2}\nonumber\\
&\approx & \frac{4}{\pi^2 N}\frac{|p|^{-\epsilon}}{\epsilon},
\label{eqG4wpdef2}
\end{eqnarray}
where we have invoked Eq.~(\ref{feynmc1}) to compute the integral
\begin{eqnarray}
\int\frac{d^n q}{(2\pi)^n}\,\frac{(\hat p+\hat
q)}{\sqrt{q^2}(p+q)^2}= \frac{\hat
p}{6\pi^2}\frac{|p|^{-\epsilon}}{\epsilon}
\end{eqnarray}
and used the $n\to 3$ asymptotic
\begin{eqnarray}
\int\limits_0^1 \frac{d\alpha}{\alpha^{n-2}}
\int\limits_0^{1/\alpha}
d\tau\left[(1-\tau)^{1-n}+(1+\tau)^{1-n}\right]
\simeq-\frac{1}{\epsilon}.
\end{eqnarray}
Notably, a singular ($1/\epsilon$) term which is present in the
gauge propagator $D^{(4)}_{\mu\nu}(q_1,q_2)$ and the origin of
which is discussed in the Appendix gets canceled in
Eq.~(\ref{eqG4wpdef2}).

Combining the four contributions (\ref{G1finres}),
(\ref{G2finres}), and (\ref{eqG4wpdef2}) we finally obtain
\begin{eqnarray}
&&\frac{1}{4}{\rm Tr}\left[\hat p \delta G(p)\right]=
\frac{1}{4}{\rm Tr}\left[\hat p\left( \delta G^{(1)}(p)+2\delta
G^{(2)}(p)+\right.\right.\nonumber\\
&&\left.\left.\delta G^{(4)}(p)\right) \right]
=\frac{32}{3\pi^2 N}\frac{|p|^{-\epsilon}}{\epsilon},\nonumber\\
\end{eqnarray}
which implies that in the momentum space Eq.~(\ref{Ginvdef})
acquires the form
\begin{equation}
G_{inv}(p)=S(p)+\delta G(p)=\frac{\hat p}{p^2} \left[1-\eta
\left(1/\epsilon-\ln |p|\right)\right],
\end{equation}
thus allowing one to read off the anomalous dimension
\begin{equation}
\eta=-\frac{32}{3\pi^2 N}. \label{eta}
\end{equation}
This result corroborates the earlier calculations performed in the
covariant and axial gauges in the framework of both the
path-integral approach of Ref.~\cite{DVK1} and the direct
perturbative expansion of Ref.~\cite{DVK2}.

\section{Gauge invariant propagator with semi-infinite strings}
In this section we set out to investigate the dependence of the
amplitude (\ref{Ginvdef}) on the choice of the contour $\Gamma$.
Specifically, we consider the contour consisting of two
(anti)parallel semi-infinite strings attached to the end points,
in which case the source current $J_\mu(z)$ in the line integral
(\ref{JA}) is given by the expression
\begin{equation}
J^\mu(z)={n}^\mu\hspace{-1.5mm}\int\limits_0^\infty
\hspace{-1.5mm}d\alpha
\left[\delta^n(z-x-{n}\alpha)\pm\delta^n(z-y\pm{n}\alpha)\right],
\label{semi-infstrings-source}
\end{equation}
where ${n}^\mu$ is a unit vector in a direction of strings. The
upper (lower) signs in (\ref{semi-infstrings-source}) correspond
to the cases of parallel and antiparallel strings, respectively.

In the former case, despite the fact that the strings do not form
a closed contour, the corresponding amplitude remains gauge
invariant, as long as all the infinitely remote points can be
compactified into a single one. This customary assumption always
holds in the perturbative sector of the gauge theory where all the
fields vanish at infinity.

The correction to the free fermion propagator has a form
\begin{eqnarray}
&&\delta G(x-y)=-\frac{1}{2}\int d^nz_1d^nz_2
J_\mu(z_1)D_{\mu\nu}(z_1-z_2)J_\nu(z_2)\nonumber\\
&&\times S(x-y)+\int d^nz_1d^nz_2 S(x-z_1)\gamma_\mu S(z_1-y)\nonumber\\
&&\times D_{\mu\nu}(z_1-z_2)J_\nu(z_2)-\int d^nz_1d^nz_2
S(x-z_1)\gamma_\mu\nonumber\\
&&\times S(z_1-z_2)\gamma_\mu S(z_2-y)D_{\mu\nu}(z_1-z_2).
\label{correction}
\end{eqnarray}
Making use of the amplitude $G(x-y)$ being explicitly gauge
invariant, we choose to compute it in the Feynman gauge where the
gauge propagator takes a particularly simple form
\begin{eqnarray}
D_{\mu\nu}(x)=\delta_{\mu\nu}\frac{A}{(x^2)^{\frac{n-1}{2}}},
\quad A=
\frac{4}{N}\frac{\mu^{3-n}\Gamma(\frac{n-1}{2})}{\pi^{\frac{n+1}{2}}}.
\end{eqnarray}
The last term in Eq.~(\ref{correction}) corresponds to the
standard fermion self-energy in the Feynman gauge
\begin{eqnarray}
\delta
G^{(3)}(\hspace{-0.4mm}x-y\hspace{-0.4mm})&\simeq&\hspace{-1mm}
\frac{4(2-n)\Gamma\left(\frac{n-1}{2}\right)}{N\pi^{\frac{n+1}{2}}(2n-3)}
\frac{|x-y|^{3-n}}{3-n} S(x-y). \label{Feynman-gauge-corr}
\end{eqnarray}
The first term in Eq.~(\ref{correction}) containing two source
currents is given by the integral
\begin{eqnarray}
&&\delta G^{(1)}(x-y)=-\frac{1}{2}\int
\hspace{-1.5mm}d^nz_1d^nz_2\,
J_\mu(z_1)D_{\mu\nu}(z_1-z_2)J_\nu(z_2)\nonumber\\
&&=-\frac{A}{2}\int\limits_0^\infty\hspace{-1.5mm}
d\alpha\hspace{-1.5mm} \int\limits_0^\infty
\hspace{-1.5mm}d\beta\, \biggr[\frac{2}{|\alpha-\beta|^{n-1}}
-\frac{1}{[(x-y+ {n}(\alpha-\beta))^2]^{\frac{n-1}{2}}}
\nonumber\\
&&-\frac{1}{[(-x+y+{n}(\alpha-\beta))^2]^{\frac{n-1}{2}}}\biggr].
\label{JDJ-infinite-xi}
\end{eqnarray}
The integral over $\beta$ is convergent for $1<n<2$. However, the
remaining integration over $\alpha$ diverges with the upper limit
$L$ which we impose as a cut-off.

Rescaling the integration variables
$\alpha\to\alpha|x-y|,\beta\to\beta|x-y|$ and introducing the
angle $\theta$ according to the relation
$\cos\theta=(x-y)\cdot{n}/|x-y|$ we rewrite
Eq.~(\ref{JDJ-infinite-xi}) as
\begin{eqnarray}
\delta
G^{(1)}(x-y\hspace{-1mm}&)&=-\frac{A|x-y|^{3-n}}{2}\hspace{-3mm}\int\limits_0^{L/|x-y|}\hspace{-4mm}
d\alpha \int\limits_0^\infty
d\beta\left[\frac{2}{|\alpha-\beta|^{n-1}}\right.\nonumber\\
&-&\left.\frac{1}{[1+2\cos\theta( \alpha-\beta)+
(\alpha-\beta)^2]^{\frac{n-1}{2}}}\right.\nonumber\\
&-&\left.\frac{1}{[1-2\cos\theta
(\alpha-\beta)+(\alpha-\beta)^2]^{\frac{n-1}{2}}}\right].
\end{eqnarray}
In the integral over $\beta$, we consider separately the intervals
from $0$ to $\alpha$ and from $\alpha$ to $\infty$. First we
compute the integral from $\alpha$ to $\infty$ which, upon
shifting the integration variable $\beta\to\beta+\alpha$, acquires
the form
\begin{eqnarray}
I_1=\int\limits_{-\infty}^\infty
d\tau\left[\frac{1}{(\tau^2)^\frac{n-1}{2}}-\frac{1}{[1-2\tau\cos\theta+
\tau^2]^{\frac{n-1}{2}}}\right].
\end{eqnarray}
After exponentiating the denominators and carrying out the
integral over $\tau$ one obtains
\begin{eqnarray}
I_1&=&\frac{\sqrt{\pi}}{\Gamma(\frac{n-1}{2})}\int\limits_0^\infty
dss^{\frac{n}{2}-2}\left[1-e^{-s\sin^2\theta}\right]\nonumber\\
&=&-\frac{\sqrt{\pi}
\Gamma(\frac{n}{2}-1)}{\Gamma(\frac{n-1}{2})}(\sin^2\theta)^{1-\frac{n}{2}}.
\end{eqnarray}
In turn, the integral over $\beta$ from $0$ to $\alpha$ takes the
form
\begin{eqnarray}
I_2&=&\int\limits_0^\alpha
\frac{d\beta}{[1+2\cos\theta(\alpha-\beta)+
(\alpha-\beta)^2]^{\frac{n-1}{2}}}\nonumber\\
&=&\alpha\int\limits_0^1\frac{d\beta}
{[1+2\alpha\beta\cos\theta+\alpha^2\beta^2]^{\frac{n-1}{2}}}\, ,
\end{eqnarray}
where we made a change of variables $\beta\to\alpha-\beta$,
followed by rescaling $\beta\to\alpha\beta$.

Next, we represent the last expresion as the difference between
the integrals taken from $0$ to $\infty$ and that from $1$ to
$\infty$
\begin{eqnarray}
&&I_2=\int\limits_0^\infty\frac{d\beta}{[\beta^2+2\beta\cos\theta
+1]^{{\frac{n-1}{2}}}}\nonumber\\
&&-(\alpha^2+2\alpha\cos\theta+1)^{1-\frac{n}{2}}\int\limits_0^\infty\frac{d\beta}
{[\beta^2+2\beta\cos\gamma+1]^{\frac{n-1}{2}}}\, , \label{I2-int}
\end{eqnarray}
where
\begin{eqnarray}
\cos\gamma=\frac{\alpha+\cos\theta}{\sqrt{\alpha^2+2\alpha\cos\theta+1}}\,.
\end{eqnarray}
The integrals in Eq.~(\ref{I2-int}) are evaluated with the use of
the formula
\begin{eqnarray}
&&\int\limits_0^\infty\frac{dx}{(x^2\pm 2x\cos\gamma+1)^\rho}=
\frac{1}{2}(\sin^2\gamma)^{\frac{1}{2}-\rho}
\frac{\sqrt{\pi}\Gamma(\rho-\frac{1}{2})}{\Gamma(\rho)}\nonumber\\
&&\mp\cos\gamma F\left(1,\rho; \frac{3}{2};\cos^2\gamma\right).
\label{key-integral}
\end{eqnarray}
Hence the sum of the two integrals reads
\begin{eqnarray}
&&I_2(\cos\theta)+I_2(-\cos\theta)=\nonumber\\
&=&(\alpha+\cos\theta)(\alpha^2+2\alpha\cos\theta+1)^{\frac{1-n}{2}}\nonumber\\
&\times& \,F\left(1,\frac{n-1}{2};\frac{3}{2};\cos^2\gamma\right)
+(\cos\theta\rightarrow-\cos\theta),
\end{eqnarray}
which can be further transformed by using the relation between the
hypergeometric functions of complementary arguments
\begin{eqnarray}
&&F(a,b;c;z)=\nonumber\\
&=&\frac{\Gamma(c)\Gamma(c-a-b)}{\Gamma(c-a)\Gamma(c-b)}
F(a,b;a+b-c+1;1-z)\nonumber\\
&+&\,(1-z)^{c-a-b}\frac{\Gamma(c)\Gamma(a+b-c)}{\Gamma(a)\Gamma(b)}\nonumber\\
&\times& \,F(c-a,c-b;c-a-b+1;1-z), \label{trasform-rule}
\end{eqnarray}
thus resulting in the expression
\begin{eqnarray}
&&I_2(\cos\theta)+I_2(-\cos\theta)=\frac{\sqrt{\pi}\Gamma(\frac{n}{2}-1)
}{2\Gamma(\frac{n-1}{2})}(\sin^2\theta)^{1-\frac{n}{2}}\nonumber\\
&&\times\left(\frac{\alpha+\cos\theta}
{|\alpha+\cos\theta|}+\frac{\alpha-\cos\theta}
{|\alpha-\cos\theta|}\right)\nonumber\\
&&+\,\frac{1}{2-n}\left[(\alpha+\cos\theta)(\alpha^2+2\alpha\cos\theta+1)^{\frac{1-n}{2}}
\right.\nonumber\\
&&\times\,\left.F\left(1,\frac{n-1}{2};\frac{n}{2};\sin^2\gamma\right)
+(\cos\theta\rightarrow-\cos\theta)\right].
\end{eqnarray}
Invoking the formula for the derivatives of the hypergeometric
function
\begin{eqnarray}
F(a+n,b;c;z)= \frac{z^{1-a}}{(a)_n}
\frac{d^n}{dz^n}[z^{a+n-1}F(a,b;c;z)]\, ,
\end{eqnarray}
we can rewrite the expression inside the square brackets in
Eq.~(58) as a total derivative. Then the integration over $\alpha$
becomes trivial and we finally get the expression for the
double-source term
\begin{eqnarray}
&&\delta G^{(1)}(x-y)=\frac{A\sqrt{\pi}\Gamma(\frac{n}{2}-1)
}{\Gamma(\frac{n-1}{2})}[L-\frac{1}{2}|x-y|\cos\theta]\nonumber\\
&&\times\,|x-y|^{2-n}(\sin^2\theta)^{1-\frac{n}{2}}-\frac{A|x-y|^{3-n}}{(2-n)(3-n)}\nonumber\\
&&\times \,F\left(1,\frac{n-3}{2};\frac{n}{2};\sin^2\theta\right),
\label{JDJ-1}
\end{eqnarray}
which behaves as
\begin{eqnarray}
\delta
G^{(1)}(x-y)\simeq\frac{4}{N\pi^2}\frac{(\mu|x-y|)^{3-n}}{3-n}\, ,
\end{eqnarray}
near $n=3$, independent of the cutoff $L$.

The second term in Eq.~(45) containing one insertion of the source
current is given by the integral
\begin{eqnarray}
\delta G^{(2)}(x-y)&=&\int\limits_0^\infty d\alpha\int
d^nz\,S(x-z)\hat{n} S(z-y)\nonumber\\
&\times&[D(z-y-\alpha{n}) -D(z-x-\alpha{n})].
\end{eqnarray}
This expression can be readily computed in the momentum space
where it reads as
\begin{eqnarray}
&&\delta
G^{(2)}(x-y)=-\int\frac{d^np}{(2\pi)^{n}}e^{-ip(x-y)}\int\limits_0^\infty
d\alpha\int\frac{d^nq}{(2\pi)^{n}}
\nonumber\\
&&\times e^{i\alpha
q{n}}[S(p)\hat{n}S(p-q)-S(p+q)\hat{n}S(p)]D(q).
\end{eqnarray}
First, we consider the integral
\begin{eqnarray}
J&=&\int\frac{d^nq}{(2\pi)^{n}} e^{i\alpha
q{n}}S(p-q)D(q)\nonumber\\
&=&\frac{8}{N}\int\frac{d^nq}{(2\pi)^{n}} e^{i\alpha q{n}}
\frac{\hat{p}-\hat{q}}{(p-q)^2}\frac{1}{\sqrt{q^2}}\, ,
\end{eqnarray}
which, upon exponentiating the denominators and integrating over
$q$, takes the form
\begin{eqnarray}
J&=&\frac{8}{N\sqrt{\pi}(4\pi)^{n/2}}\int\limits_0^\infty
ds\int\limits_0^\infty\frac{dt}{\sqrt{t}}\frac{1}{(s+t)^{\frac{n}{2}+1}}\nonumber\\
&\times&\,e^{-p^2\frac{st}{s+t}+i\alpha\frac{s}{s+t}p{n}
-\frac{\alpha^2}{4(s+t)}}
\left(\hat{p}t-\frac{i\alpha\hat{n}}{2}\right).
\end{eqnarray}
Thus Eq.~(63) can be written as
\begin{eqnarray}
&&\delta
G^{(2)}(x-y)=-\frac{8}{N\sqrt{\pi}(4\pi)^{n/2}}\int\frac{d^np}{(2\pi)^{n}}
e^{-ip(x-y)}\nonumber\\
&&\times\int\limits_0^\infty d\alpha \int\limits_0^\infty
ds\int\limits_0^\infty\frac{dt}{\sqrt{t}}
\frac{1}{(s+t)^{\frac{n}{2}+1}}
e^{-p^2\frac{st}{s+t}-\frac{\alpha^2}{4(s+t)}}
\left[e^{i\alpha\frac{s}{s+t}p n}\right.\nonumber\\
&&\times\left.S(p)\hat{n}\left(\hat{p}t-\frac{i\alpha{\hat
n}}{2}\right)- e^{-i\alpha\frac{s}{s+t}p n}
\left(\hat{p}t+\frac{i\alpha{\hat
n}}{2}\right){\hat n}S(p)\right].\nonumber\\
\end{eqnarray}
After inserting into the integrand the identity $\int_0^\infty
d\rho\delta(\rho-s-t)=1$ and rescaling the variables $s\to
s\rho,t\to t\rho$ one can readily perform the integration over
$s$.

The integration over $\rho$ results in the table integral
\begin{eqnarray}
\int\limits_0^\infty
dxx^{\alpha-1}e^{-px-q/x}=2\left(\frac{q}{p}\right)^{\alpha/2}K_\alpha(2\sqrt{pq}),
\end{eqnarray}
thus yielding
\begin{eqnarray}
&&\delta
G^{(2)}(x-y)=-\frac{32i}{N\sqrt{\pi}(4\pi)^{n/2}}\int\frac{d^np}{(2\pi)^{n}}
e^{-ip(x-y)}\nonumber\\
&&\times (2|p|)^{\frac{n-3}{2}}\int\limits_0^\infty
d\alpha\alpha^{\frac{3-n}{2}}
\int\limits_0^1dt[t(1-t)]^{\frac{n-3}{4}}\left[\sqrt{1-t}\right.\nonumber\\
&&\times\left.\sin(\alpha tp{n})
S(p)\hat{n}\hat{p}\,K_{\frac{3-n}{2}}
(\alpha|p|\sqrt{t(1-t)})\right.\nonumber\\
&&-\left.|p|S(p)\sqrt{t}\cos(\alpha tp{n})
K_{\frac{1-n}{2}}(\alpha|p|\sqrt{t(1-t)})\right].
\end{eqnarray}
The remaining integrals over $\alpha$ are given by the formulas
$6.699.3$ and $6.699.4$ from the Integral Tables \cite{Gradstein}.

Thus, we arrive at the formula
\begin{eqnarray}
&&\delta
G^{(2)}(x-y)=-\frac{i2^{5-n}}{N\pi^{\frac{n+1}{2}}}\int\hspace{-1.5mm}\frac{d^np}{(2\pi)^{n}}
e^{-ip(x-y)}|p|^{n-4}\hspace{-1.5mm}\nonumber\\
&&\times\int\limits_0^1\hspace{-1.5mm}dt
t^{\frac{n-3}{2}}(1-t)^{\frac{n-4}{2}}\left[S(p)\hat{n}\hat{p}
\cdot\frac{pn}{|p|}\Gamma\left(\frac{5-n}{2}\right)\right.\nonumber\\
&&\times\left.F\left(1,\frac{5-n}{2};\frac{3}{2};
-\frac{t(pn)^2}{(1-t)p^2}\right)-\frac{1}{2}|p|S(p)\right.\nonumber\\
&&\left.\times\,\Gamma\left(\frac{3-n}{2}\right)F\left(1,\frac{3-n}{2};\frac{1}{2};
-\frac{t(pn)^2}{(1-t)p^2}\right)\right],
\end{eqnarray}
where the integration over $t$ can be performed by changing the
variable $t=u/(1+u)$ and comparing the result with the integral
representation for the hypergeometric function ${}_3F_2$ of a
certain argument.

However, one can notice that at $n\to3$ the main contribution
stems from the second term in the square brackets
\begin{eqnarray}
\delta
&&G^{(2)}(x-y)\simeq\frac{2\Gamma\left(n-\frac{3}{2}\right)\Gamma\left(\frac{3-n}{2}
\right)\mu^{3-n}}{N\pi^{n+\frac{1}{2}}\Gamma\left(\frac{5-n}{2}\right)}
\frac{\hat{x}-\hat{y}}{[(x-y)^2]^{n-\frac{3}{2}}}\nonumber\\
&&=\frac{8\Gamma\left(n-\frac{3}{2}\right)}
{N\pi^{\frac{n+1}{2}}\Gamma\left(\frac{n}{2}\right)}\frac{(\mu|x-y|)^{3-n}}{3-n}S(x-y),
\end{eqnarray}
where we restored the dependence on the dimensionful parameter
$\mu$ and also used
\begin{eqnarray}
S(x)=\frac{\Gamma(n/2)}{2\pi^{n/2}}\frac{\hat{x}}{(x^2)^{n/2}}.
\end{eqnarray}
Combining Eqs.~(47), (61), and (70) together, we find the overall
correction to the fermion propagator
\begin{eqnarray}
G(x-y)\approx\left[1+\frac{32}{3\pi^2N}\frac{(\mu|x-y|)^{3-n}}{3-n}\right]S(x-y),
\end{eqnarray}
from which one can read off the anomalous dimension.

Remarkably, the latter appears to be still given by
Eq.~(\ref{eta}), as in the case of the original ''short-cut''
contour studied in the previous Section.

Furthermore, a similar calculation shows that the negative
anomalous dimension Eq.~(\ref{eta}) also pertains to the case of
the parallel strings which corresponds to choosing the upper sign
in Eq.~(\ref{semi-infstrings-source}).

Taken at their face value, these observations suggest that the
gauge invariant amplitude Eq.~(\ref{Ginvdef}) may even be largely
independent of the choice of the contour $\Gamma$.

\section{Summary}
In this work, we carried out a direct calculation of the
previously conjectured form of the physical electron propagator in
such effective $QED$-like models as the theory of the pseudogap
phase of the cuprates. In contrast to the earlier work, we
performed our calculations in the reliable radial gauge and
confirmed the result (\ref{eta}) obtained in
Refs.\cite{DVK1,DVK2}.

In the course of our analysis, we also investigated the dependence
of the amplitude (2) on the choice of the contour $\Gamma$ by
considering the case of two (anti)parallel semi-infinite strings
attached to the end points. The corresponding gauge invariant
amplitude is given by Eq.~(\ref{Ginvdef}) with the current
Eq.~(\ref{semi-infstrings-source}) entering the line integral
(\ref{JA}). Remarkably, the algebraic behavior (1) controlled by
the same negative anomalous exponent (\ref{eta}) appears to be
valid for these functions as well.

In addition to the possible dependence on the choice of the
contour $\Gamma$ (or a lack thereof), the anomalous dimension may
strongly depend on the massless fermion amplitude in question. For
instance, when computed in one of the covariant gauges, the
amplitude
\begin{equation}
G_\xi(x-y)={\langle 0|\psi(x)\exp{\left[i(\xi-1)\int
\limits_y^xdz^\mu\, A_\mu(z)\right]}\bar\psi(y)|0 \rangle\over
{\langle 0|\exp{\left[i\xi\int \limits_y^xdz^\mu\,
A_\mu(z)\right]}|0 \rangle}} \label{Brown}
\end{equation}
exhibits a positive anomalous dimension
\begin{equation}
\eta_\xi={16\over 3\pi^2N}(3\xi-2) \label{eta-xi}
\end{equation}
for any $\xi>2/3$ \cite{DVK2}, including the case of $\xi=1$ which
has been claimed \cite{Franz2} to provide an identical
representation of the original function $G_0(x)$ given by
Eq.~(\ref{Ginvdef}).

However, for any $\xi\neq 0$ the amplitude $G_\xi(x)$ given by
(\ref{Brown}) is not truly gauge-invariant, and, in particular,
its anomalous dimension computed in a non-covariant gauge may
differ from Eq.~(\ref{eta-xi}). (for an extended discussion of
this subtle point, see \cite{DVK2}). For instance, when computed
in the radial gauge applied in this paper the anomalous dimension
of the function $G_\xi(x)$ turns out to be independent of $\xi$
and is given by Eq.~(\ref{eta}).

This makes it clearly impossible to substitute any of the
surrogate amplitudes $G_\xi(x)$ with $\xi>2/3$ (e.g., $G_1(x)$, as
in Ref.~\cite{Franz2}) for the original one, $G_0(x)$, which is
the only truly gauge-invariant member of the family of functions
(\ref{Brown}).

Evidently, the negative anomalous dimension manifested by the
function (2) contradicts the anticipated behavior of a viable
candidate to the role of the physical electron propagator, since
in all of the previously discussed effective $QED_3$-like models
the repulsive electron interactions are expected to result in
further suppression, rather than enhancement, of any amplitude
describing propagation of physical electrons.

In particular, the algebraic decay of the gauge-invariant fermion
amplitude (1) would only result in the sought-after Luttinger-like
(stronger-than-linear) vanishing density of states
$\nu(\epsilon)\sim |\epsilon|^{1+\eta}$ if $\eta$ were positive.
By the same token, a pseudogap theory can only reconcile with the
experimentally established absence of well-defined nodal
quasiparticles if a branch-cut singularity of the electron
propagator that occurs at $p^2_\mu=0$ appears to be weaker (not
stronger!) than a simple pole.

We defer a further discussion of the construction of the physical
electron propagator until future work (see, however,
Refs.\cite{DVK1,DVK2} for an alternate form which demonstrates a
faster-than-algebraic decay, thus further diminishing the chances
that the conjecture about the Luttinger-like behavior in $QED_3$
may still be "right, albeit for a wrong reason").

Instead, we suggest that the negative anomalous dimension
(\ref{eta}) of the heuristically chosen gauge-invariant amplitude
(2) may pertain not so much to the physical electron propagator
$per$ $se$, but rather to the vertex corrections which also
control the behavior of various gauge-invariant two-particle
amplitudes ("susceptibilities").

In this regard, we quote the earlier result of Ref.~\cite{GHR}
obtained for the susceptibility associated with the four-fermion
scalar vertex
\begin{equation}
<0|{\overline \psi}(x)\psi(x){\overline \psi}(y)\psi(y)|0> \propto
{1\over |x-y|^{4-(64/3\pi^2N)}} \label{chi}
\end{equation}
which features a negative anomalous dimension $2\eta$. In the
context of the $QED_3$ theory of the pseudogap phase of the
cuprates, the formula (\ref{chi}) describes the divergence of the
 staggered spin susceptibility
at the antiferromagnetic ordering vector ${\vec Q}=(\pi,\pi)$
\cite{Wen2}.

We emphasize that one encounters the above problem with the
unphysical (slower than $\propto 1/x^{2}$) decay of the amplitude
(2) only in the massless case, while for a finite fermion mass
this function decays as $\propto e^{-m|x|}$.

In the case of the $QED_3$ theory of the pseudogap phase, it has
been argued that one may indeed expect a dynamical mass generation
corresponding to the intrinsic instability towards a spin and/or
charge density wave ordering \cite{Herbut1,Franz2}.

The question remains, though, as to whether or not the chiral
symmetry breaking can at all occur for the physical number of
fermion flavors ($N=2$). Even in the fully Lorentz-invariant
situation there exist some analytical \cite{app} and numerical
\cite{hands} results which suggest the upper bound $N_{cr}<2$ for
the critical number of flavors below which the chiral symmetry
gets broken.

In the (non-Lorentz-invariant) $QED_3$ theory of the pseudogap phase
of Refs.\cite{Lee,Wen1,Franz1,Ye1,Herbut1},
the role of the strong spatial anisotropy of the quasiparticle
dispersion
and, in particular, its effect on a possible universality  (or a lack thereof)
of the critical value of $N_{cr}$ still remain to be ascertained
(see \cite{anisotropy} for a discussion of the weakly anisotropic case).

It is worth mentioning, however, that in the extreme
non-Lorentz-invariant limit of the $QED_3$-like theory describing
the problem of layered graphite the estimated value of $N_{cr}$
was found to be even lower than in the original Lorentz-invariant
case \cite{graphite}.

We conclude by stressing that the problem of constructing the true
physical electron propagator in the effective massless $QED$-like
theories still remains unresolved. Nevertheless, our calculation
confirms once and for all that the naive ansatz (2) is not up to
the job, thereby eliminating the current basis for the theoretical
predictions of the Luttinger-like behavior in the underdoped
cuprates \cite{Wen1,Franz1,Ye1}.

It is, however, conceivable that, while being unrelated to the
actual behavior of the electron propagator, the negative anomalous
dimension (\ref{eta}) manifests the same properties of the gauge
invariant vertex corrections as those exhibited by the physically
relevant two-fermion amplitudes such as Eq.~(\ref{chi}).

\section*{Acknowledgments}
This research was supported in part by the National Science
Foundation under Grants No. PHY-0070986 (V. P. G.) and DMR-0071362
(D. V. K.) and by the SCOPES-projects 7~IP~062607 and
7UKPJ062150.00/1 of Swiss NSF (V. P. G.). One of the authors (D.
V. K.) achnowledges hospitality at Aspen Center for Physics and
NORDITA (Copenhagen, Denmark) where part of this work was carried
out.

\appendix
\section{Fock-Schwinger photon propagator}
\label{FS-propagator}
In this Appendix we demonstrate that the line integral in
Eq.~(\ref{Ginvdef}) vanishes in the so-called radial or
Fock-Schwinger (FS) gauge. We also derive the expression for the
photon propagator in this gauge.

The FS gauge is defined as
\begin{equation}
(x-x_0)^\mu A_\mu(x,x_0)=0. \label{Fock-gauge}
\end{equation}
In contrast to such widely used gauges as the Landau $\partial^\mu
A_\mu(x)=0$, the Coulomb $\partial_iA_i(x)=0 (i=1,2)$ and the
axial $n_\mu A_\mu(x)=0$ ones, the FS gauge may break the
translational invariance because of the presence of a fixed point
$x_0$. However, an important advantage of the Fock gauge is the
explicit relation between the potential $A_\mu(x,x_0)$ and the
field strength $F_{\mu\nu}$
\begin{equation}
A_\mu(x,x_0)=\int\limits_0^1d\alpha\,\alpha(x-x_0)^\nu
F_{\nu\mu}(\alpha(x-x_0)+x_0,x_0). \label{A-through-F}
\end{equation}
In order to derive Eq.~(\ref{A-through-F}) we differentiate
Eq.~(\ref{Fock-gauge})
\begin{equation}
A_\mu(x,x_0)+(x-x_0)^\nu\partial_\mu A_\nu(x,x_0)=0,
\label{differentiated-FS}
\end{equation}
and then use $F_{\mu\nu}=\partial_\mu A_\nu-\partial_\nu A_\mu$,
to write
\begin{equation}
A_\mu(x,x_0)+(x-x_0)^\nu(F_{\mu\nu}(x,x_0)+\partial_\nu
A_\mu(x,x_0))=0.
\end{equation}
Upon changing the variable $x\to\alpha(x-x_0)+x_0$ the last
expression turns into
\begin{eqnarray}
&&\frac{d}{d\alpha}[\alpha A_\mu(\alpha(x-x_0)+x_0,x_0)]\nonumber\\
&&=\alpha(x-x_0)^\nu F_{\nu\mu}(\alpha(x-x_0)+x_0,x_0).
\end{eqnarray}
Integrating over $\alpha$ and using the boundary condition
$A_\mu(x_0,x_0)=0$ (see, Eq.~(\ref{differentiated-FS})) which
assumes the regularity of $A_\mu(x,x_0)$ at $x=x_0$, we arrive at
Eq.~(\ref{A-through-F}). Note that the boundary condition
$A_\mu(x_0,x_0)=0$ is essential for eliminating a residual gauge
freedom which remains even after imposing the gauge condition
(\ref{Fock-gauge}).

Indeed, in addition to the solution (\ref{A-through-F})
Eq.~(\ref{Fock-gauge}) can be satisfied by any function
\begin{equation}
A^0_\mu(x,x_0)=\partial^x_\mu f(x-x_0),
\end{equation}
where $f$ is an arbitrary homogeneous function of $x-x_0$ of zero
degree. Any such function would necessarily be singular at
$x=x_0$, though. Hence, the regularity condition at $x-x_0$ can be
used to fix the residual gauge freedom in (\ref{Fock-gauge}).

Under the translation $U_a^{-1}F_{\mu\nu}(x)U_a=F_{\mu\nu}(x-a)$
the solution (A2) transforms as
\begin{equation}
U_a^{-1}A_\mu(x,x_0)U_a=A_\mu^\prime(x,x_0)=A_\mu(x-a,x_0-a).
\end{equation}
When expressed in terms of the center mass $X=(x+y)/2$ and the
relative $\bar{x}=x-y$ coordinates the line integral in
Eq.~(\ref{Ginvdef}) takes the following form
\begin{eqnarray}
I(\bar{x},X;x_0)=\int\limits_{y}^x dz_\mu A^\mu(z)\nonumber\\
=(x-y)^\mu\int\limits_0^1d\alpha A_\mu(\alpha(x-y)+y,x_0)\nonumber\\
=\bar{x}^\mu\int\limits_{-1/2}^{1/2}d\alpha A_\mu
(\alpha\bar{x}+X,x_0). \label{function-I}
\end{eqnarray}
Under translations, Eq.~(\ref{function-I}) transforms according to
the rule: $I(\bar{x},X;x_0)=I(\bar{x},X-a;x_0-a)$.

We can further restrict the gauge condition (A1) by choosing the
fixed point $x_0$ at the center of mass, {\it i.e.}, $x_0=X$
\begin{equation}
({x}-X)^\mu A_\mu({x},X)=0 \label{Fock-gauge-special}
\end{equation}
(hereafter, we simplify the notation $A_\mu({x},x_0=X)\equiv
A_\mu({x})$).

It can be readily seen that in the gauge
(\ref{Fock-gauge-special}) the line integral vanishes, {\it i.e.},
$I(\bar{x},X;X)=0$. Indeed, from Eqs.~(\ref{function-I}) and
(\ref{A-through-F}) we obtain
\begin{eqnarray}
& &I(\bar{x},X;X)=\bar{x}^\mu\hspace{-2mm}\int\limits_{-1/2}^{1/2}
d\alpha  A_\mu(\alpha\bar{x}+X)\nonumber\\
&=&\bar{x}^\mu\bar{x}^\nu\int\limits_{-1/2}^{1/2}
d\alpha\int\limits_0^1d\beta\beta
F_{\nu\mu}(\alpha\beta\bar{x}+X)=0
\end{eqnarray}
due to the antisymmetry of $F_{\nu\mu}$.

Furthermore, performing a translation with $a=X$ we can cast the
gauge condition (\ref{Fock-gauge-special}) in the form
\begin{equation}
({x}-X)^\mu A_\mu({x}-X)=0,
\end{equation}
which is identical to
\begin{equation}
{x}^\mu A_\mu({x})=0. \label{our-gauge}
\end{equation}
Next, we derive the photon propagator in the gauge
(\ref{our-gauge}) where Eq.~(A2) reads as 
\begin{equation}
A_\mu(x)=\int\limits_0^1d\alpha\,\alpha x^\nu F_{\nu\mu}(\alpha
x).
\end{equation}
Thus, we find
\begin{eqnarray}
& &D^{FS}_{\mu\nu}(x,y)=\langle0|TA_\mu(x)A_\nu(y)|0\rangle\nonumber\\
&=&\int\limits_0^1d\alpha d\beta\,\alpha\beta x^\sigma
y^\rho\langle0|TF_{\sigma\mu}(\alpha x)F_{\rho\nu}(\beta
y)|0\rangle.
\end{eqnarray}
Since the field strength $F_{\mu\nu}$ is a gauge invariant
quantity, the correlator
$\langle0|TF_{\sigma\mu}(x)F_{\rho\nu}(y)|0\rangle$ can be
calculated in any gauge, including, e.g., the Feynman one where it
becomes
\begin{eqnarray}
&&x^\sigma y^\rho\langle0|TF_{\sigma\mu}(x)F_{\rho\nu}(y)|0\rangle
=x^\sigma
y^\rho\left(\delta_{\mu\nu}\partial^x_\sigma\partial^y_\rho-\delta_{\mu\rho}
\partial^x_\sigma\partial^y_\nu\right.\nonumber\\
&&-\left.\delta_{\sigma\nu}\partial^x_\mu\partial^y_\rho
+\delta_{\sigma\rho}\partial^x_\mu\partial^y_\nu\right)D(x-y)\nonumber\\
&&\equiv H_{\mu\nu}(x,y)D(x-y).
\end{eqnarray}
Here $D(x)$ is the photon propagator in the Feynman gauge
\begin{equation}
D(x)=\frac{A}{(x^2)^{\frac{n-1}{2}}},\quad
A=\frac{4}{N}\frac{\mu^{3-n}\Gamma(\frac{n-1}{2})}{\pi^\frac{n+1}{2}}.
\end{equation}
With the use of the relation $x_\mu\partial_\mu=|x|\partial_{|x|}$
the operator $H_{\mu\nu}(x,y)$ can be written in the form
\begin{eqnarray}
H_{\mu\nu}(x,y)&=&\delta_{\mu\nu}\partial_{|x|}\partial_{|y|}|x||y|-\partial^x_\mu
x_\nu \partial_{|y|}|y|\nonumber\\
&-&\partial^y_\nu y_\mu\partial_{|x|}|x|+
\partial^x_\mu\partial^y_\nu x\cdot y.
\end{eqnarray}
Now, making use of the identity
\begin{eqnarray}
&&\partial_{|x|}\int\limits_0^1d\alpha\,|x|f(\alpha
x)=\partial_{|x|}\int\limits_0^1
d\alpha\,|x|f(\alpha|x|\hat{ x})\nonumber\\
&&=\partial_{|x|}\int\limits_0^{|x|}d\beta f(\beta\hat{ x})=f(x),
\end{eqnarray}
where $\hat{ x}=x/|x|$ is the unit vector, we obtain
\begin{eqnarray}
&&D_{\mu\nu}(x,y)=\langle0|TA_\mu(x)A_\nu(y)|0\rangle\nonumber\\
&&=\int\limits_0^1d\alpha
d\beta\, H_{\mu\nu}(\alpha x,\beta y)D(\alpha x-\beta y)\nonumber\\
&&=H_{\mu\nu}(x,y)\int
\limits_0^1d\alpha d\beta D((\alpha x-\beta y)^2)=\delta_{\mu\nu}D(x-y)\nonumber\\
&&-\int\limits_0^1d\alpha\,\partial^x_\mu x_\nu D(\alpha x-y)
-\int\limits_0^1d\beta\,\partial^y_\nu y_\mu D(x-\beta y)\nonumber\\
&&+\int\limits_0^1d\alpha d\beta\,
\partial^x_\mu\partial^y_\nu x\cdot yD(\alpha x-\beta y)
\label{propagator-in-FSgauge}
\end{eqnarray}
(cf. Refs.\cite{Skachkov,Leupold}).

While the first term in (\ref{propagator-in-FSgauge}) depends
solely on $\bar{x}$, the others have a more complicated
dependence, hence the photon propagator computed in an arbitrary
FS gauge is not necessarily translationally invariant. Moreover,
the last term in Eq.~(\ref{propagator-in-FSgauge}) displays a
divergence at $n=3$ which forces one to use the method of
dimensional regularization when computing various quantities. As
pointed out in Ref.~\cite{Leupold} (also, see
Eq.~(\ref{eqG4wpdef2})), the divergence of the free FS gauge
propagator at $n=3$ dimensions is, in fact, necessary for
obtaining the correct results.

Finally, in order to return to the gauge
(\ref{Fock-gauge-special}) we replace $x\to x-X, y\to y-X$ in
(\ref{propagator-in-FSgauge}), thus obtaining the FS gauge
propagator $D^{FS}_{\mu\nu}(x,y)$ as a sum of the four terms (here
the arguments $x$ and $y$ are unrelated to the end points in the
line integral)
\begin{equation}
D^{(1)}_{\mu\nu}(x,y)\equiv\delta_{\mu\nu} D(x-y),\label{D1def}
\end{equation}
\begin{equation}
D^{(2)}_{\mu\nu}(x,y)\equiv -\int\limits_0^1 d\alpha\partial_\mu^x
(x-X)_\nu D(\alpha x-y+(1-\alpha)X),
\end{equation}
\begin{equation}
D^{(3)}_{\mu\nu}(x,y)\equiv -\int\limits_0^1 d\beta\partial_\nu^y
(y-X)_\mu D(x-\beta y-(1-\beta)X),
\end{equation}
\begin{eqnarray}
D^{(4)}_{\mu\nu}(x,y)&\equiv& \int\limits_0^1 d\alpha
d\beta\partial_\mu^x \partial_\nu^y
(x-X)\cdot (y-X) D(\alpha x-\beta y\nonumber\\
& &-(\alpha-\beta)X). \label{D4def}
\end{eqnarray}

\end{document}